\documentclass[aps,floatfix,prd,preprint]{revtex4-1}
\usepackage{graphicx,bm,amsmath,color,tabularx}
\usepackage[utf8]{inputenc} 
\def\*{^{(*)}}
\def\be{\begin{equation}}
\def\ee{\end{equation}}

\def\Lc{\Lambda_c}

\def\S{\Sigma_c}
\def\jp{J/\psi}
\def\jpp{\jp\, p}
\def\D{\bar D}

\def\>{\big>}
\def\<{\big<}
\def\|{\big\vert}

\def\etal{\textit{et~al.~}}

\begin{document}
\title{Experimental constraints on the properties of $P_c$ states}
\author{T.\,J.\,Burns}
\affiliation{Department of Physics, Swansea University, Singleton Park, Swansea, SA2 8PP, UK.}
\author{E.\,S.\,Swanson}
\affiliation{Department of Physics and Astronomy, University of Pittsburgh, Pittsburgh, PA 15260, USA.}

\begin{abstract}

Using new experimental data on photoproduction and $\Lambda_b^0$ decays, we derive  constraints on the properties of the LHCb $P_c$ states. We conclude that $P_c(4312)$, $P_c(4380)$ and $P_c(4440)$ can be described as $\S\D$, $\S^*\D$ and $\S\D^*$ molecules, but that $P_c(4457)$ does not fit into the same picture. Based on the apparent absence of additional partner states, and the striking disparity between $\Lc^+\D^0$ and $\Lc^+\D^{*0}$ decays, we conclude that $P_c(4440)$ has $3/2^-$ quantum numbers. Using heavy-quark symmetry we predict large experimental signals for $P_c$ states in  $\eta_c p K^-$, $\Lc^+\D^{*0}K^-$, and $\S\*\D K^-$. We also argue that current experimental data on photoproduction is almost at the level of sensitivity required to observe $P_c$ states.
\end{abstract}

\maketitle 

\section{Introduction}
\label{Sec:introduction}

The LHCb $P_c$ states have so far been observed only in the $\jpp$ spectrum of $\Lambda_b\to J/\psi\, p \,K^-$ decays \cite{Aaij:2015tga,Aaij:2016phn,Aaij:2019vzc}, but experimental data from other processes can also usefully constrain their properties. In this paper we explore the consequences of some recent experimental observations whose implications for the phenomenology of the $P_c$ states have not been recognised in the literature. In particular, we consider: tighter constraints on photoproduction and $\jpp$ decays from the $\jp$\textit{-007} experiment, measurements of $\Lambda_b^0$ decays to  $\Lc^+\D^{0}K^-$, $\Lc^+\D^{*0}K^-$ and $\eta_c p K^-$ (including limits on $P_c$ fit fractions), and the apparent absence of $P_c$ partners with higher mass.

In the molecular scenario, we find that these new experimental results imply constraints on the quantum numbers of $P_c$ states, and give striking predictions for prominent production channels that can be tested in experiment. Our observations follow directly from experiment, with minimal theoretical assumptions. At most, we rely only on symmetry principles which are well-justified theoretically and empirically: heavy-quark symmetry, and the dominance of colour-favoured processes in weak transitions.

We begin with an overview of molecular scenarios for the $P_c$ states (Sec.~\ref{sec:scenarios}), and then review the recent experimental data which forms the basis of our analysis (Sec.~\ref{sec:expt}). We then consider the resulting phenomenology associated with various decay modes, namely $\jpp$~(Sec.~\ref{sec:jpp}), $\eta_cp$~(Sec.~\ref{sec:ep}), $\Lc^+\D^0$~(Sec.~\ref{sec:ld}), $\Lc\D\*\pi$~(Sec.~\ref{sec:tb}), $\Lc^+\D^{*0}$~(Sec.~\ref{sec:lds}) and $\S\*\D$~(Sec.~\ref{sec:sd}). We finally consider the implications of the apparent absence of heavier partners to the $P_c$ states~(Sec.~\ref{sec:partners}), and summarise our main results~(Sec.~\ref{sec:conclusions}).

\section{Scenarios}
\label{sec:scenarios}

In the discovery paper~\cite{Aaij:2015tga}, an amplitude analysis of $\Lambda_b\to J/\psi\, p \,K^-$ identified two states decaying to $\jpp$: the broad $P_c(4380)$ and narrower $P_c(4450)$. A model-independent analysis~\cite{Aaij:2016phn} confirmed the importance of including these contributions in adequately describing the data. A subsequent analysis, with an order of magnitude more data~\cite{Aaij:2019vzc}, discovered an additional new state $P_c(4312)$, and resolved the initial $P_c(4450)$ into two distinct, narrow peaks, $P_c(4440)$ and $P_c(4457)$. 

The original $P_c(4380)$ did not feature in the newer analysis, which was sensitive only the narrow features in the spectrum. Because its discovery relied on an amplitude model which has been superseded, the existence or otherwise of  $P_c(4380)$  remains to be determined experimentally, as explained in the note in the appendix of ref.~\cite{Aaij:2019vzc}. In our discussion we will refer to the ``$P_c(4380)$'', although we will not insist that its measured properties are consistent with those extracted from the original analysis, which is regarded as obsolete.

Because of their proximity to thresholds, the $P_c$ structures are widely interpreted as molecular states whose wavefunctions are dominated by S-wave combinations of $\S\*\D\*$ constituents \cite{Wu:2010vk,Roca:2015dva,He:2015cea,Chen:2015loa,Karliner:2015ina,Ortega:2016syt,Shimizu:2016rrd,Yamaguchi:2016ote,Yamaguchi:2017zmn,Shimizu:2017xrg,Shimizu:2018ran,Du:2019pij,Gutsche:2019mkg,Liu:2019zvb,Valderrama:2019chc,
Sakai:2019qph,Guo:2019kdc,He:2019ify,Liu:2019tjn,Chen:2019asm,Burns:2019iih,He:2019rva,Xiao:2019aya,Peng:2020xrf,Xu:2020gjl,Yalikun:2021bfm,Du:2021fmf}. So $P_c(4312)$ is a $1/2^-$ $\S\D$ state, while $P_c(4380)$ is $3/2^-$ $\S^*\D$ state, albeit with significantly smaller width that the state observed in the original LHCb analysis. The $P_c(4440)$ and $P_c(4457)$ are both $\S\D^*$ states, and can be assigned to either $1/2^-$ and $3/2^-$, or $3/2^-$ and $1/2^-$, respectively. We summarise these assignments in the first two rows of Table~\ref{tab:scenarios}, where the Scenarios ``A'' and ``B'' correspond to those of ref.~\cite{Liu:2019tjn}. 
\begin{table}
    \centering
    \begin{tabularx}{\textwidth}{XXXXX}
\hline
&$P_c(4312)$&$P_c(4380)$&$P_c(4440)$&$P_c(4457)$\\
\hline
Scenario A & $1/2^-~\S\D$&$3/2^-~\S^*\D$& $1/2^-~\S\D^*$& $3/2^-~\S\D^*$\\
Scenario B & $1/2^-~\S\D$&$3/2^-~\S^*\D$& $3/2^-~\S\D^*$& $1/2^-~\S\D^*$\\
Scenario C & $1/2^-~\S\D$&$3/2^-~\S^*\D$& $3/2^-~\S\D^*$& see text\\
\hline
    \end{tabularx}
    \caption{Quantum numbers and dominant degrees of freedom in the various scenarios.}
    \label{tab:scenarios}
\end{table}

\begin{figure}[t]
    \centering
    \includegraphics[width=0.8\textwidth]{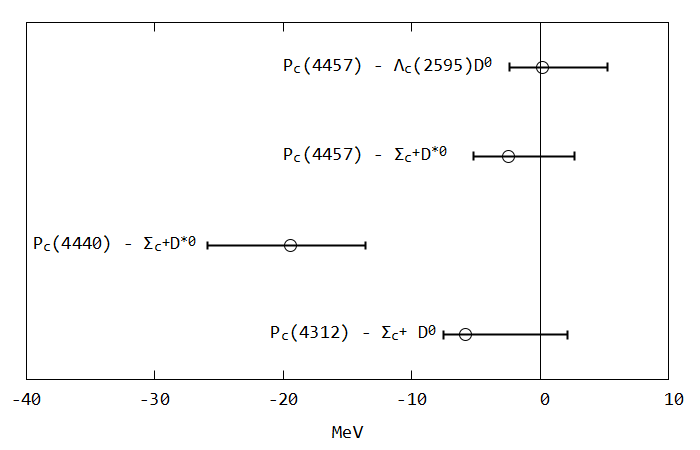}
    \caption{The binding energies of $P_c(4312)$, $P_c(4440)$ and $P_c(4457)$ with respect to their nearest thresholds.  The $P_c(4380)$ is not shown, due to the large uncertainty in its mass (which overlaps considerably with the $\S^*\D$ threshold).}
    \label{fig:binding}
\end{figure}

The implicit assumption in most models is that the states are bound with respect to the nearest $\S\*\D\*$ threshold, but experimentally,  this has only been established for $P_c(4440)$. In Fig.~\ref{fig:binding} we show the binding energies of $P_c(4312)$, $P_c(4440)$ and $P_c(4457)$ with respect to their nearest thresholds. While $P_c(4440)$ is unambiguously bound with respect to $\S^+\D^{*0}$, the masses of the other states are consistent, within uncertainties, with the thresholds. 

The case of $P_c(4457)$ is particularly noteworthy. Its mass is consistent not only with the the $\S^+\D^{*0}$ threshold but also, more strikingly, with $\Lc(2595)\D$ threshold: the difference in the central values is just 0.2~MeV. This naturally suggests that for $P_c(4457)$, the $\S\D^*$ bound state scenario is not the only possibility. In a forthcoming paper we explore a number of viable alternatives \cite{inprep}. Given its proximity to threshold, the most prosaic option is that it is a threshold cusp arising from $\S\D^*\to\jpp$ or $\Lc(2595)\D\to \jpp$, or some combination of the two. Indeed the cusp scenario is favoured in the analysis of ref.~\cite{Kuang:2020bnk}. Alternatively, the peak could arise as a triangle singularity in $\Lc(2595)\D$ or, as in our previous paper \cite{Burns:2019iih}, as a genuine resonance, but with $1/2^+$ quantum numbers and $\Lc(2595)\D$ degrees of freedom. We  find that all of these scenarios give a very good fit to the $\Lambda_b\to J/\psi\, p \,K^-$ data \cite{inprep}.

It turns out to be very liberating to abandon the hypothesis that $P_c(4457)$ is a bound $\S\D^*$ state, and not only because the alternative scenarios give a good fit to data. As we argue below, there are some phenomenological problems with the usual modelling, all of which can be traced to the assumption that $P_c(4457)$ is a bound $\S\D^*$ state. By abandoning this assumption we avoid these problems.

Our alternative Scenario ``C'', also shown in the Table~\ref{tab:scenarios}, differs from the other two in no longer assuming that $P_c(4457)$ is a $\S\D^*$ bound state. The assignments of the other states are consistent with Scenario B; the rationale for preferring $3/2^-$ quantum numbers for $P_c(4440)$ is explained below. Note that our Scenario C actually incorporates several possibilities, corresponding to different explanations for $P_c(4457)$.

Of course other scenarios are also discussed in the literature. In particular, we highlight refs.~\cite{Nakamura:2021qvy,Nakamura:2021dix} in which non-resonant interpretations of $P_c(4312)$ have been discussed.

\section{Experimental data}
\label{sec:expt}

In Table~\ref{tab:expt} we show the experimental data which forms the basis of much of our discussion. For each of four possible $\Lambda_b$ decays (indicated in the top row), we show the three-body branching fraction ($\mathcal B$) and, where data are available, a measure ($\mathcal R$) of the contribution of $P_c$ states in the fit, defined below. The data in the last three columns are quite new, and their significance has not yet been appreciated in literature.

The three-body branching fractions $\mathcal B$ for $\Lambda_b^0 \to \jpp K^-$ and $\Lambda_b^0 \to \eta_c pK^-$ are from  LHCb \cite{Aaij:2015fea,Aaij:2020mlx,ParticleDataGroup:2020ssz}, while those of $\Lambda_b^0 \to \Lc^+\D^0K^-$ and $\Lambda_b^0 \to \Lc^+\D^{*0}K^-$ are from Stahl~\cite{Stahl:2018eme}, combining the measured ratios
\begin{align}
    \frac{\mathcal B(\Lambda_b^0 \to \Lc^+\D^0K^-)}{\mathcal B(\Lambda_b^0 \to \Lc^+D_s^-)}&=(14.04\pm0.58\pm0.33\pm 0.45)\%,\\
    \frac{\mathcal B(\Lambda_b^0 \to \Lc^+\D^{*0}K^-)}{\mathcal B(\Lambda_b^0 \to \Lc^+D_s^-)}&=(43.5\pm 1.4^{+1.2}_{-0.8}\pm 1.4)\%,
\end{align}
from LHCb data, with the Particle Data Group (PDG) value for the denominators \cite{ParticleDataGroup:2020ssz}
\begin{equation}
    \mathcal B(\Lambda_b^0 \to \Lc^+D_s^-)=(1.1\pm 0.1)\%.
\end{equation}
\begin{table}
    \centering
    \begin{tabularx}{\textwidth}{XXXXX}
\hline
&$\Lambda_b^0 \to \jpp K^-$&$\Lambda_b^0 \to \eta_c pK^-$&$\Lambda_b^0 \to \Lc^+\D^0K^-$&$\Lambda_b^0 \to \Lc^+\D^{*0}K^-$\\
\hline
$\mathcal B~(/ 10^{-4})$& $3.2^{+0.6}_{-0.5}$ &  $1.06\pm 0.26$&$15\pm 2$&$48\pm 5$\\
\hline
$\mathcal R~(/10^{-2})$&&&$1/2^-\qquad3/2^-$&\\
\cline{4-4}
\quad$P_c(4312)$&$0.30 \pm 0.07 {}^{+0.34}_{-0.09}$&$<24$&$<0.4$&\\
\quad$P_c(4440)$&$1.11 \pm 0.33 {}^{+0.22}_{-0.10}$&&$<0.6\phantom{0}\quad<0.45$&\\
\quad$P_c(4457)$&$0.53 \pm 0.16 {}^{+0.15}_{-0.13}$&&$<0.45\quad<0.8$&\\
\hline
    \end{tabularx}
    \caption{Experimental data on the three-body branching fractions $\mathcal B$, and $P_c$ fit fractions $\mathcal R$, in $\Lambda_b^0$ decays. The sources of experimental data are described in the text. All upper limits are at 95\% confidence level. As shown, in the case of $\Lambda_b^0 \to \Lc^+\D^0K^-$, the limits on fit fractions depend on the assumed quantum numbers ($1/2^-$ or $3/2^-$) of the $P_c$ states.}
    \label{tab:expt}
\end{table}

The quantity $\mathcal R$ is a measure of the contribution of the $P_c$ states in the fit, although the experimental procedures in how this number is computed differ. For $\Lambda_b^0 \to \jpp K^-$ and $\Lambda_b^0 \to \eta_c pK^-$, the numbers are from refs.~\cite{Aaij:2019vzc,Aaij:2020mlx}, where they are described as ``relative contributions'', and are identified in terms of branching fractions as 
\begin{equation}
\mathcal R(\jpp) = \frac{
{\mathcal B}(\Lambda_b^0 \to P_c^+ K^-) \, 
{\mathcal B}(P_c^+ \to J/\psi \,p) } {
{\mathcal B}(\Lambda_b^0 \to J/\psi \, p\, K^-) },\label{eq:ffjp}
\end{equation}
and
\begin{equation}
{\mathcal R (\eta_c p)} = \frac{
{\mathcal B}(\Lambda_b^0 \to P_c^+ K^-) \, 
{\mathcal B}(P_c^+ \to \eta_c p) } {
{\mathcal B}(\Lambda_b^0 \to \eta_c p K^-) }.\label{eq:ffep}
\end{equation}
The numbers for $\Lambda_b^0 \to \Lc^+\D^0K^-$, from the  amplitude analysis of Piucci~ \cite{Piucci:2019vsk}, are ``fit fractions'', defined in the usual way: the ratio, for a single resonance versus the full amplitude,  of the phase space integrals of the square of the matrix element. (In the original $P_c$ discovery paper \cite{Aaij:2015tga}, the contributions of $P_c(4380)$ and $P_c(4450)$ were also quantified in terms of fit fractions.) We adopt the usual interpretation of the fit fraction, which is widespread in the experimental literature and the PDG~\cite{ParticleDataGroup:2020ssz}, in terms of branching fractions, namely
\begin{equation}
{\mathcal R (\Lc^+\D^0)} = \frac{
{\mathcal B}(\Lambda_b^0 \to P_c^+ K^-) \, 
{\mathcal B}(P_c^+ \to \Lc^+\D^0) } {
{\mathcal B}(\Lambda_b^0 \to \Lc^+\D^0K^-) }.\label{eq:ffld}
\end{equation}
Note that the limits on the fit fractions depend on the assumed quantum numbers ($1/2^-$ or $3/2^-$) of the $P_c$ states, as shown in the table.

The difference between the ``relative contributions'' of refs.~\cite{Aaij:2019vzc,Aaij:2020mlx} and the ``fit fractions'' of refs.~\cite{Aaij:2015tga,Piucci:2019vsk} relates to the treatment of interference terms, and  is discussed in ref.~\cite{Aaij:2019vzc}. There is however no difference in the way the numbers are interpreted in terms of branching fractions, as shown in equations~\eqref{eq:ffjp}, \eqref{eq:ffep} and~\eqref{eq:ffld}. Hence we will use the generic term ``fit fraction'', and the same symbol ${\mathcal R}$, for both quantities.

We remark, however, that for both the ``relative contributions'' and ``fit fractions'', the standard interpretation of these experimental numbers in terms of branching fractions has an implicit assumption, which is not necessarily justified. The factorisation of the numerator into a product of branching fractions is legitimate if the $P_c$ signal is due to a resonance, but not, for example, if it arises from a kinematical singularity or threshold cusp, in which case it is not possible (conceptually or mathematically) to separate the production ($\Lambda_b^0 \to P_c^+ K^-$) and decay (such as $P_c^+\to\jpp$). We discussed this point previously in relation to the $X(2900)$ states~\cite{Burns:2020xne}, for which several competing interpretations are possible~\cite{Burns:2020epm}.

Much of our analysis below involves manipulations of measured $\mathcal R$ values, and in particular it relies on the standard factorisation of the branching fractions in the numerators. In our preferred Scenario C, we can only assume this factorisation for $P_c(4312/4380/4440)$, which we are treating as $\S\*\D\*$ resonances. We cannot assume factorisation for $P_c(4457)$ because it is not justified for some of the scenarios applicable to this state (such as threshold cusp and triangle singularity); hence in much of the discussion below we will not consider $P_c(4457)$.

In computing the fit fractions ${\mathcal R (\Lc^+\D^0)}$,  the masses and widths of $P_c$ states, as determined in previous experiments, are taken as inputs \cite{Piucci:2019vsk}. For this reason we only quote the results for $P_c(4312/4440/4457)$, rather than those of the precursor states $P_c(4380/4450)$, whose mass and width measurements are considered to be obsolete~\cite{Aaij:2019vzc}.

There is a caveat on the limits on $\Lambda_b^0 \to \Lc^+\D^0K^-$ fit fractions that appear in the table. These were obtained by testing separately for the presence of a single state in the data -- either $P_c(4312)$, $P_c(4440)$ or $P_c(4457)$ -- rather than all three simultaneously. (The latter approach was found to be numerically unfeasible.) 

While this suggests a degree of caution is warranted in interpreting the figures, it is reassuring to consider the results obtained for the precursor states $P_c(4380)$ and $P_c(4450)$, where both types of analyses were performed. The model in which both states were included simultaneously gives comparable, but somewhat tighter, upper limits on the branching fractions, compared to the model in which the presence of each state was tested separately. If the same applies in the case of $P_c(4312/4440/4457)$, the upper limits quoted in the table would be conservative.

The discussion so far has concentrated on $P_c$ states in 3-body $\Lambda_b$ decays. Another possibility, much discussed in the literature, is to search for the states in photoproduction~\cite{Wang:2015jsa,Karliner2015a,Kubarovsky:2016whd,Ali:2019lzf,Blin:2016dlf,Joosten2021}. Currently there is no evidence for $P_c$ states in $\gamma p\to P_c\to\jpp$, but the measured upper limits, which have recently become tighter, have interesting implications.

In particular, recall that the absence of the states in $\gamma p\to P_c\to\jpp$ implies upper limits, via vector meson dominance, on the $P_c\to\jpp$ branching fractions. After the initial discovery of the original LHCb states $P_c(4450)$ and $P_c(4380)$, analysis of  photoproduction data available at the time implied an upper limit on the $\jpp$ branching fractions at the percent level~\cite{Wang:2015jsa}. More recently, the  GlueX experiment~\cite{Ali:2019lzf} obtained similar upper limits from their new data, using a variant of the JPAC model for the amplitude~\cite{Blin:2016dlf}. The GlueX results assume $3/2^-$ quantum numbers for all the states, however choosing a different assignment is not expected to change the results drastically. The results are 
\begin{equation}
\mathcal B (P_c^+\to J/\psi\, p)<(2.3\%\div 4.6\%),
\end{equation}
depending on which of $P_c(4312/4440/4457)$ is being considered. 

These stringent upper limits have recently become even tighter: preliminary results from another Jefferson Lab experiment, $J/\psi$-\textit{007}, indicate that the $\jpp$ upper limits are smaller than the GlueX limits by a factor of approximately 10 \cite{Joosten2021}. 

We may summarise the situation, rather roughly, as
\begin{equation}
    {\mathcal B}(P_c^+\to J/\psi\, p)<(\textrm{a few})\times10^{-3}.\label{eq:jplimit}
\end{equation}
This is adequate for present purposes, as our arguments below only require estimates of the scales of the branching fractions, rather than their precise values. (We are also limited in not having an estimate of the systematic uncertainty associated with the assumed $3/2^-$ quantum numbers in the JPAC/GlueX analysis.)

\section{$\jpp$ decays }
\label{sec:jpp}

We now come to the main thrust of the paper, exploring the implications of the experimental data discussed in the previous section. We begin with $\jpp$ decays, and note that the stringent upper limit~\eqref{eq:jplimit} is in conflict with the vast majority of models, in some cases by orders of magnitude \cite{Wu:2010jy,Wu:2010vk,Roca:2015dva,Shen:2016tzq,Lu:2016nnt,Ortega:2016syt,Lin:2017mtz,Huang:2018wed,Xiao:2019mvs,Wang:2019spc,Xu:2019zme,Eides:2019tgv,Xiao:2019mvs,Xiao:2020frg,Ruangyoo:2021aoi}. Our own estimate \cite{Burns:2019iih}, for $P_c(4440)$, is one of very few that is consistent with~\eqref{eq:jplimit}. Similarly, Dong~\etal~\cite{Dong:2020nwk} obtain a $P_c(4312)$ branching fraction which is near to the upper limit~\eqref{eq:jplimit}. Eides~\etal~\cite{Eides:2018lqg} computed $\jpp$ branching fractions in both the molecular and hadrocharmonium pictures, noting a very different pattern in decays; their results for the molecular scenario, but not the hadrocharmonium scenario, are consistent with~\eqref{eq:jplimit}.

Superficially, the tight upper limit on $\jpp$ decays may appear surprising, considering that the $P_c$ states were of course discovered as prominent peaks in the $\jpp$ spectrum. This illustrates the point that in quantifying the prominence of a structure in the 2-body spectrum of a 3-body decay, the most relevant quantity is not the 2-body decay branching fraction itself, but the fit fraction $\mathcal R$. Recalling equation~\eqref{eq:ffjp}, the $P_c$ fit fractions are a  measure of the fraction of all $\jpp K^-$ events that are produced via $\Lambda_b^0 \to P_c^+ K^-,P_c^+ \to J/\psi \,p$. Obviously, the fit fractions therefore depend not only on the properties of the $P_c$ states themselves (the production and decay branching fractions in the numerator), but also on the sum total of all other production mechanisms resulting in the $\jpp K^-$ final state (the denominator). It turns out that tree-level contributions to $\jpp\,K^-$ are suppressed, resulting in enhanced $P_c$ fit fractions. 

\begin{figure}
    \centering
    \includegraphics[width=\textwidth]{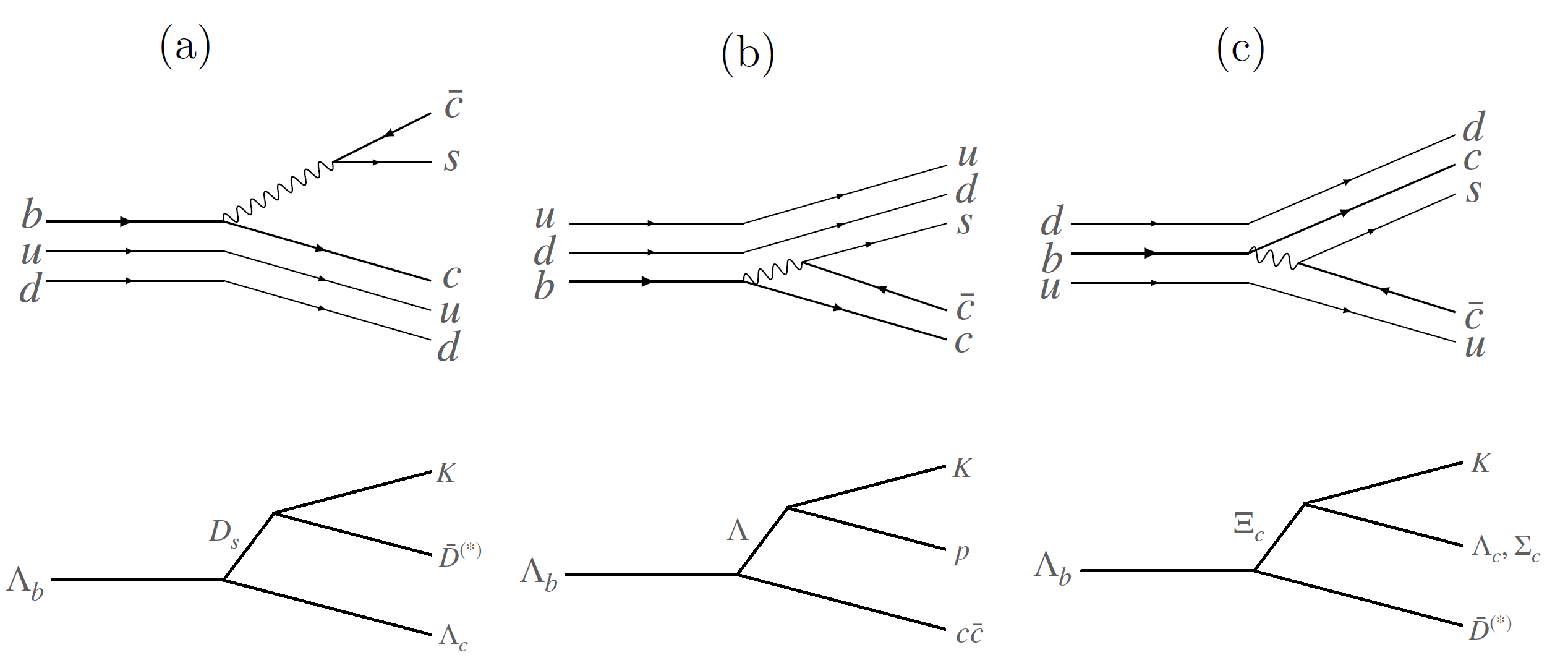}
    \caption{Tree level contributions to three-body $\Lambda_b^0$ decays. Diagram (a) is colour-enhanced, whereas diagrams (b) and (c) are colour-suppressed.}
    \label{fig:tree}
\end{figure}

The suppression is explained in Fig.~\ref{fig:tree}. There are three possible flavour topologies at the weak vertex (top panel), and for each, there is a corresponding three-body tree-level diagram that produces a kaon (bottom panel). The weak vertex in
Fig.~\ref{fig:tree}(a) is colour-enhanced, whereas those of Fig.~\ref{fig:tree}(b) and~\ref{fig:tree}(c) are colour-suppressed \cite{Burns:2019iih}. On this basis we expect, very roughly,
\begin{align}
    {\mathcal B}(\Lambda_b^0 \to \Lc^+\D^{(*)0}K^-)&>>{\mathcal B(\Lambda_b^0 \to c\bar c\, p\, K^-)}
    \approx
    {\mathcal B(\Lambda_b^0 \to \S\*\,\D\*\,K^-)},\label{eq:colsup}
\end{align} which is consistent with the numbers in Table~\ref{tab:expt}.

We conclude that the $P_c$ structures in $\Lambda_b^0 \to \jpp\, K^-$ are prominent (namely, have large fit fractions) in part because of the suppression of tree-level $\Lambda_b^0 \to \jpp\, K^-$ decays. Intriguingly, the situation is similar for several other exotic states observed in 3-body weak decays. For example,  $\chi_{c1}(3872)$ [formerly $X(3872)$] has comparatively small branching fraction to  $\jp\pi^+\pi^-$ and $\jp\omega$ \cite{ParticleDataGroup:2020ssz}, but was discovered and extensively studied in $B\to \jp\pi^+\pi^- K$ and $B\to \jp\omega K$, transitions for which the tree-level diagrams are also colour-suppressed. (Note that $\chi_{c1}(3872)$ has much larger branching fraction to $D^0\D^{*0}$, for example, but the tree-level diagram $D^0\D^{*0}K$ is colour-enhanced, and much larger.) In ref.~\cite{Burns:2020xne} we noted a similar mechanism enhancing the $X(2900)$ signal in $B^+\to D^+X,X\to D^-K^+$ \cite{Aaij:2020hon,Aaij:2020ypa}. 

\begin{figure}
	\includegraphics[width=0.9\textwidth]{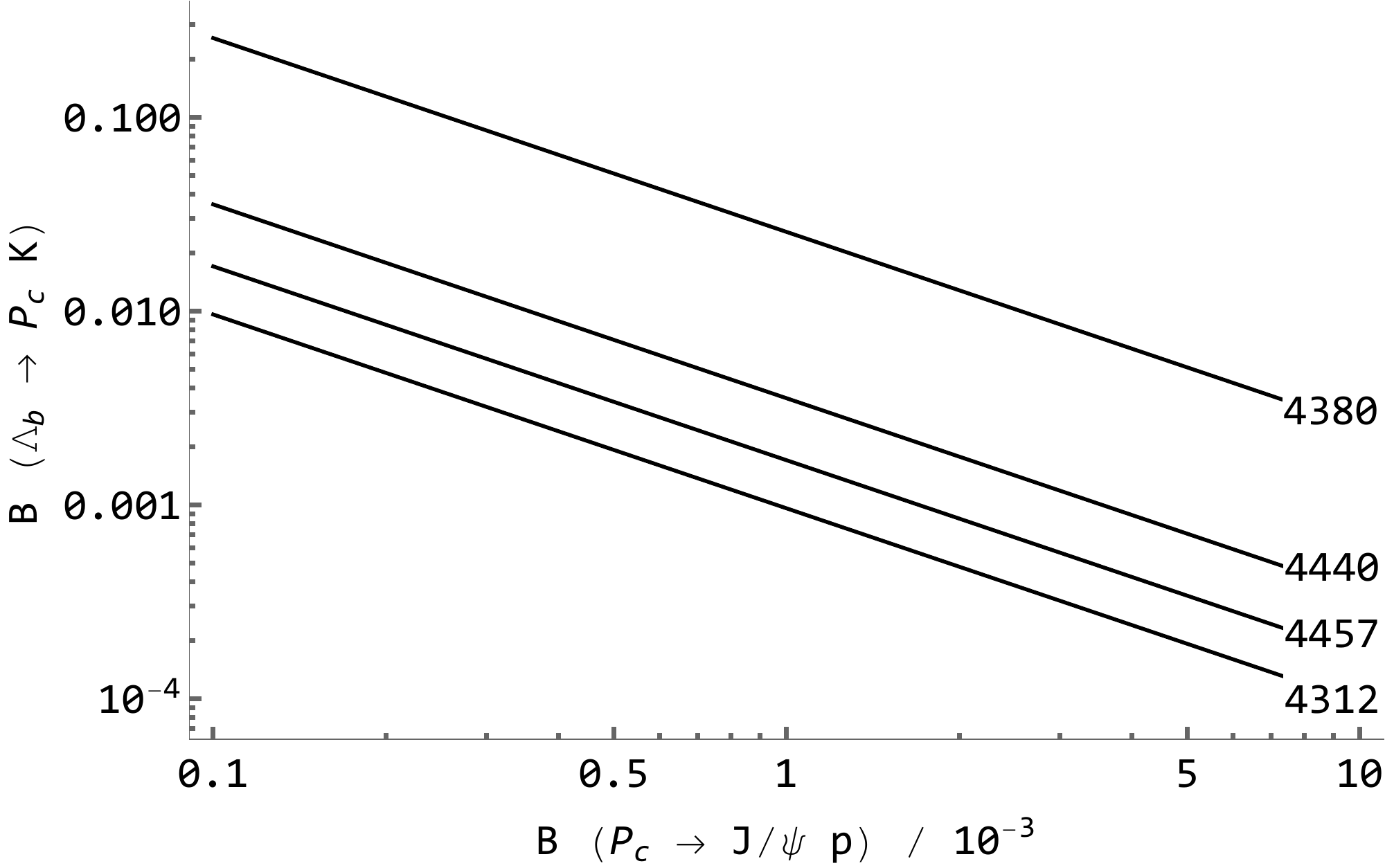}
	\caption{The branching fractions $\mathcal B (\Lambda_b^0\to P_c^+ K^-)$ as a function of $\mathcal B (P_c^+\to J/\psi\, p)$, obtained as described in the text.}
	\label{fig:bfs}
\end{figure}

We return now to the tighter upper limit~\eqref{eq:jplimit} on $\jpp$ decays, and note that it  has striking implications for $P_c$ production in $\Lambda_b^0$ decays. Upper limits on ${\mathcal B}(P_c^+ \to J/\psi \,p)$ imply lower limits on ${\mathcal B}(\Lambda_b^0 \to P_c^+\,K^-)$~\cite{Cao:2019kst}. Following our previous paper~\cite{Burns:2019iih}, in Fig.~\ref{fig:bfs} we plot ${\mathcal B}(\Lambda_b^0 \to P_c^+\,K^-) $ as a function of ${\mathcal B}(P_c^+ \to J/\psi \,p)$, using equation~\eqref{eq:ffjp}, and the experimental values (Table~\ref{tab:expt}) for ${\mathcal B}(\Lambda_b^0 \to J/\psi \, p\, K^-)$ and ${\mathcal R}(\jpp)$. We have also included $P_c(4380)$ in the plot, using the measured value~\cite{Aaij:2015tga}
\begin{align}
\mathcal{R}(\jpp)&=(8.4 \pm 0.7 \pm 4.2)\%,\label{eq:ff04}
\end{align}
although as noted below, this part of the plot should be interpreted with some caution. Note that in the plot we are showing the central values only; we do not include error bars, because the discussion below concerns the overall scale of the branching fractions, not their precise values.

Combining Fig.~\ref{fig:bfs} with equation~\eqref{eq:jplimit} we conclude that $\mathcal B (\Lambda_b^0\to P_c^+ K^-)$ is at least ${\mathcal O}(10^{-3})$ for $P_c(4312/4440/4457)$, and ${\mathcal O}(10^{-2})$ for $P_c(4380)$. (These are larger by a factor of 10 than our previous estimates~\cite{Burns:2019iih}, because the limit on the $\jpp$ branching fraction is now smaller by a factor of 10.) Strikingly, these numbers are comparable to the largest measured two-body branching fraction for $\Lambda_b^0$ decays~\cite{ParticleDataGroup:2020ssz}
\begin{equation}
    {\mathcal B}(\Lambda_b^0\to \Lc^+D_s^-)=(1.1\pm 0.1)\%.
\end{equation}

The comparison is quite awkward because naively one would expect the production of multiquark states (whether molecular or compact) to be suppressed compared to that of conventional hadrons, and indeed that is what is observed in other sectors. The closest analogy is with $\chi_{c1}(3872)$, for which the measured branching fraction~\cite{ParticleDataGroup:2020ssz},
\begin{equation}
    {\mathcal B}(B^+\to \chi_{c1}(3872)K^+)=(2.1\pm 0.7)\times 10^{-4},
\end{equation}
is around a factor of fifty smaller than the corresponding branching fractions for conventional hadrons $B^+\to D_s\*\D\*$. Indeed, if anything, we would expect an even stronger relative suppression for the $P_c$ states because, unlike in the case of $\chi_{c1}(3872)$, their dominant wavefunction components cannot be produced in colour-favoured processes~\cite{Burns:2019iih}.

With this comparison in mind, the production branching fractions for $P_c(4380)$ seem implausibly large, but as noted previously, the numbers in this case should be treated with caution: we have used a measured fit fraction which, along with other properties of this state, are now considered to be obsolete~\cite{Aaij:2019vzc}.

But even the less dramatic numbers for $P_c(4312/4440/4457)$ present a challenge for models, and of course, the challenge becomes more acute as the limits on branching fractions ${\mathcal B}(P_c^+\to\jpp)$ become tighter. This suggests that ${\mathcal B}(P_c^+\to\jpp)$ cannot be much less than the current upper bound~\eqref{eq:jplimit}, otherwise $\mathcal B (\Lambda_b^0\to P_c^+ K^-)$ will become implausibly large. Indirectly, it suggests that the sensitivity required to observe $P_c$ states in photoproduction is not much more than that of the $J/\psi${\textit{-007}} experiment.

Note that the production branching fractions implied by the above analysis are orders of magnitude larger than those predicted by the effective Lagrangian approach of ref.~\cite{Wu:2019rog}.

\section{$\eta_c p$ decays}
\label{sec:ep}

For $1/2^-$ states the S-wave decays to $\jpp$ and $\eta_c p$ are related by heavy-quark symmetry~\cite{Voloshin:2019aut,Sakai:2019qph}. In our Scenario C there is a single $1/2^-$ state, the $P_c(4312)$ with $\S\D$ constituents. Ignoring differences due to phase space, the relation among branching fractions is
\begin{equation}
    \frac{{\mathcal B}[P_c(4312)\to \eta_c p]}{{\mathcal B}[P_c(4312)\to \jpp]}=3.
\end{equation}
We conclude, by comparison to  equation~\eqref{eq:jplimit}, that 
\begin{equation}
    {\mathcal B}[P_c(4312)\to\eta_cp]<{\mathcal O} (1\%)\label{eq:eplimit}
\end{equation}

Nevertheless, the experimental prospects in $\Lambda_b\to \eta_c p K^-$ decays are encouraging: as noted previously, the relevant quantity is not the branching fraction ${\mathcal B}(P_c\to \jpp)$, but the fit fraction ${\mathcal R}(\eta_cp)$. From equations~\eqref{eq:ffjp} and \eqref{eq:ffep} we find
\begin{equation}
    \frac{\mathcal R (\eta_c p)}{\mathcal R(\jpp)}=3~\frac{{\mathcal B}(\Lambda_b^0 \to J/\psi \, p\, K^-)}{{\mathcal B}(\Lambda_b^0 \to \eta_c p K^-)}
\end{equation}
and thus, using the numbers from Table~\ref{tab:expt}, we predict a substantial fit fraction for $P_c(4312)$:
\begin{equation}
    {\mathcal R}(\eta_c p)=(2.7^{+4.3}_{-2.0})\%
\end{equation}
This is large in comparison to the fit fractions for the discovery mode $\Lambda_b^0 \to J/\psi \, p\, K^-$, suggesting the experimental prospects are encouraging. Note that the prediction is consistent with the experimental upper limit in Table~\ref{tab:expt}.

As for the other states, for $P_c(4457)$ we cannot make any prediction for $\eta_cp$ decays in our Scenario C, without further modelling. (As noted previously, there are several viable scenarios for this state.) However for $P_c(4380)$ or $P_c(4440)$ we anticipate, on general grounds, negligible $\eta_cp$ signals, as they both have $3/2^-$ quantum numbers and thus do not couple to $\eta_c p$ in S-wave.

This is quite different from Scenarios A and B, in which one of $P_c(4440)$ or $P_c(4457)$ is a $\S\D^*$ bound state with $1/2^-$ quantum numbers, and thus couples to $\eta_c p$ in S-wave. Drawing on the results of ref.~\cite{Voloshin:2019aut}, we find
\begin{equation}
    \frac{\mathcal R (\eta_c p)}{\mathcal R(\jpp)}=\frac{3}{25}~\frac{{\mathcal B}(\Lambda_b^0 \to J/\psi \, p\, K^-)}{{\mathcal B}(\Lambda_b^0 \to \eta_c p K^-)}.
\end{equation}
This implies, in comparison to equation~\eqref{eq:jplimit}, that if $P_c(4440)$ or $P_c(4457)$ is a $1/2^-$ state,
\begin{equation}
    {\mathcal B}[P_c(4440/4457)\to\eta_cp]<{\mathcal O} (10^{-3}).
\end{equation}
In terms of fit fractions, in  Scenario A we predict 
\begin{equation}
    {\mathcal R}(\eta_c p)=\left(0.40^{+0.31}_{-0.29} \right)\%
\end{equation}
for $P_c(4440)$ and negligible $\eta_cp$ for $P_c(4457)$, whereas for Scenario B we predict
\begin{equation}
    {\mathcal R}(\eta_c p)=\left(0.19 \pm 0.16\right)\%
    \end{equation}  
for $P_c(4457)$, and negligible $\eta_cp$ for $P_c(4440)$. The ${\mathcal R}(\eta_cp)$ fit fractions in these scenarios are comparable to the measured ${\mathcal R}(\jpp)$ fit fractions in Table~\ref{tab:expt}, suggesting such measurements may be within reach in future analyses. Confronting these predictions with data can discriminate among Scenarios A, B and C.

\section{$\Lc^+\D^0$ decays}
\label{sec:ld}

We now turn to $\Lc^+\D^0$ decays, which in many models are expected to be prominent channels. As we will see, this is not borne out by experimental data.

For a given $P_c$ state, combining equations \eqref{eq:ffjp} and \eqref{eq:ffld} yields a relation between the $\Lc^+\D^0$ and $\jpp$ branching fractions
\begin{equation}
    \frac{{\mathcal B}(P_c^+\to\Lc^+\D^0)}{{\mathcal B}(P_c^+ \to J/\psi \,p) }=\frac{{\mathcal R}(\Lc^+\D^0)}{{\mathcal R}(\jpp)}\frac{{\mathcal B(\Lambda_b^0 \to \Lc^+\D^0K^-)}}{{\mathcal B(\Lambda_b^0 \to J/\psi \, p\, K^-)} }.
\end{equation}
Taking the experimental data from Table~\ref{tab:expt} we obtain \begin{equation}
    \frac{{\mathcal B}(P_c^+\to\Lc^+\D^0)}{{\mathcal B}(P_c^+ \to J/\psi \,p) }<(1.9\div 7.1),
\end{equation}
depending on which of the states $P_c(4312/4440/4457)$ is considered, and its assumed quantum numbers. In combination with the photoproduction upper limit~\eqref{eq:jplimit}, we arrive at the surprising result  \begin{equation}{\mathcal B}(P_c^+\to\Lc^+\D^0)\lesssim \mathcal O(1\%).\label{eq:ld1}\end{equation}
We will show later that this limit argues in favour of  Scenario~C.

\section{$\Lc\D\*\pi$ decays}
\label{sec:tb}
In previous sections we observed that the $P_c$ branching fractions to $\jpp$, $\eta_c p$ and $\Lc^+\D^0$ are tiny. This implies that their measured decay widths,
\begin{align}
\Gamma[P_c(4312)]&=9.8\pm 2.7^{+3.7}_{-4.5}~\textrm{MeV},\label{eq:G1}\\
\Gamma[P_c(4440)]&=20.6\pm4.9^{+8.7}_{-10.1}~\textrm{MeV},\label{eq:G2}\\
\Gamma[P_c(4457)]&=6.4\pm 2.0^{+5.7}_{-1.9}~\textrm{MeV},\label{eq:G3}
\end{align}
must be dominated by other modes. One possibility is three-body decays which, in the molecular scenario, would arise via the decay of a constituent hadron. Given that $\D$ is stable and $\D^*$ has negligible decay width, molecular three-body decays are presumably dominated by $\S\to\Lc\pi$ or $\S^*\to\Lc\pi$~\cite{Burns:2015dwa}. Following refs.~\cite{Swanson:2006st,Voloshin:2019aut,Burns:2019iih}, we assume that the three-body width is determined by the width of the $\S\*$ constituent, resulting in the following predicted partial widths
\begin{align}
    \Gamma[P_c(4312)\to \Lc^+\D\pi]&=1.9~\textrm{MeV}\label{eq:G4}\\
    \Gamma[P_c(4380)\to \Lc^+\D\pi]&=15~\textrm{MeV}\label{eq:G5}\\
    \Gamma[P_c(4440/4457)\to \Lc^+\D^{*}\pi]&=1.9~\label{eq:G6}\textrm{MeV}
\end{align}
Evidently these three-body decays cannot account entirely for the measured $P_c$ decay widths. Given the tiny $\jpp$ branching fractions,  we also disregard  more esoteric possibilities like $\chi_{c0}p$ and $\jp N\pi$.

\section{$\Lc^+\D^{*0}$ decays}
\label{sec:lds}

We have established that the partial widths of $P_c$ states to $\jpp$, $\eta_c p$ and $\Lc^+\D^0$ are tiny, and that three-body modes, while not negligible, cannot account for the measured decay widths. The rest of the decay width must be accounted for by other modes. We assume that decays to exclusively light-flavoured hadrons are small; this is because of the OZI rule, and the apparent absence of prominent light-flavoured decays in other hidden-charm states.

In the case of $P_c(4312)$ we conclude, by a process of elimination, that the dominant decay must be $\Lc^+\D^{*0}$, as this is the only two-body mode with hidden charm which is kinematically accessible. Ignoring the small contributions from $\jpp$, $\eta_c p$ and $\Lc^+\D^0$, we estimate the $\Lc^+\D^{*0}$ partial width by taking the difference of equations~\eqref{eq:G1} and \eqref{eq:G4},
\begin{equation}
    \Gamma[P_c^+(4312)\to\Lc^+\D^{*0}]=2.7\div12.5~\textrm{MeV},\label{eq:ld3}
\end{equation}
where we have combined the statistical and systematic uncertainties in~\eqref{eq:G1} in quadrature. We have not accounted for the uncertainties in equation~\eqref{eq:G4}, because of the difficulty in quantifying the systematic uncertainty on the underlying assumption, namely that the three-body width is equal to the width of the free constituents. Most of our arguments below relate to the scale of the experimental numbers, not their precise values, so we do not expect the conclusions to be compromised by the unquantified uncertainties.

One may question the validity of predicting prominent $\Lc^+\D^{*0}$ modes on the basis of a process of elimination. However we also note that our conclusion is supported by model calculations which confirm the prominence of $\Lc^+\D^{*0}$ decays in molecular models~\cite{Ortega:2016syt,Shimizu:2017xrg,Lin:2017mtz,Huang:2018wed,He:2019rva,Dong:2020nwk}. 

From our estimate~\eqref{eq:ld3}, the corresponding branching fraction
\begin{equation}
    {\mathcal B}[P_c^+(4312)\to\Lc^+\D^{*0}]=59\%\div 87\%,\label{eq:ld2}
\end{equation}
is enormous in comparison to the closely related $\Lc^+\D^0$ mode, equation~\eqref{eq:ld1}. The challenge for models is to explain the striking disparity between these two modes which would naively be expected to be comparable. We will see that this issue argues in favour of our Scenario~C.

Heavy quark symmetry implies relations among all couplings of the type $\S\*\D\*\to \Lc^+\D^{0(*)}$. In Table~\ref{tab:hqs} we show the relative matrix elements for S-wave transitions obtained assuming heavy-quark symmetry, taken from the potentials in ref.~\cite{Du:2021fmf}. Note that the relative matrix elements apply not only to effective field theory models, where the S-wave potentials correspond to contact terms that are fit to data, but also to models where the transition is due to the central potential from one-pion exchange (since such models satisfy heavy-quark symmetry).

\begin{table}
    \centering
    \begin{tabularx}{\textwidth}{XXXXX}
\hline
&$\S\D$&$\S^*\D$&$\S\D^*$&$\S\D^*$\\
&$1/2^-$&$3/2^-$&$1/2^-$&$3/2^-$\\
\hline
$\Lc\D$&0&&$\sqrt 3$&\\
$\Lc\D^*$&$\sqrt 3$&$-\sqrt 3$&$-2$&1\\
\hline
Scenario A &$P_c(4312)$&$P_c(4380)$&$P_c(4440)$&$P_c(4457)$\\
Scenario B &$P_c(4312)$&$P_c(4380)$&$P_c(4457)$&$P_c(4440)$\\
Scenario C &$P_c(4312)$&$P_c(4380)$&&$P_c(4440)$\\
\hline
    \end{tabularx}
    \caption{Relative matrix elements for S-wave transitions $\S\*\D\*\to\Lc\D\*$, according to heavy-quark symmetry. The values are extracted from ref.~\cite{Du:2021fmf}, but apply not only to contact terms in effective field theory, but also to the central potentials due to one-pion exchange. The lower part of the table identifies the various scenarios.}
    \label{tab:hqs}
\end{table}

The first thing to notice is that for $P_c(4312)$, modelled as a $1/2^-$ $\S\D$  state, the striking disparity in the magnitudes of $\Lc\D$ and $\Lc\D^*$ has a natural explanation, and confirms a selection rule predicted by Voloshin~\cite{Voloshin:2019aut}. Although conservation of angular momentum allows for decays to both $\Lc\D$ (in S-wave) and $ \Lc\D^*$ (S-wave and D-wave), the $\Lc\D$ decay is forbidden by heavy-quark symmetry. In this sense the experimental data are nicely consistent with the molecular model for $P_c(4312)$. (The suppression of the $\Lc\D$ mode is also a feature of the chiral constituent quark model~\cite{Dong:2020nwk}, and the chromomagnetic pentaquark model~\cite{Weng:2019ynv}, but not the model of ref.~\cite{He:2019rva}.)

The tight upper limit~\eqref{eq:ld1} on $\Lc\D$ decays also has a natural explanation for $3/2^-$ states, both $P_c(4380)$ ($\S^*\D$) and $P_c(4440/4457)$ ($\S\D^*$). In these cases the $\Lc\D$ decay would be D-wave, hence suppressed compared to the S-wave decay $P_c(4312)\to\Lc\D^*$.

On the other hand, for a $1/2^-$ $\S\D^*$ state, the tight upper limit on $\Lc^+\D^0$ is a problem. The previous argument relies on the assumption (which is very natural) that D-wave decays are suppressed compared to S-wave decays. In that case, the $\Lc\D$ and $\Lc\D^*$  partial widths for the various $P_c$ states are dominated by the S-wave matrix elements, so we can estimate the relative strengths using the numbers in Table~\ref{tab:hqs}. We notice in particular that for $1/2^-$, the $\S\D^*\to\Lc^+\D^0$ matrix element is identical to $\S\D\to \Lc^+\D^{*0}$, whose scale is set by equation~\eqref{eq:ld3}. This is  a problem for Scenarios A and B, because it implies that if either of $P_c(4440)$ or $P_c(4457)$ were a $1/2^-$ $\S\D^*$ state, then its $\Lc^+\D^0$ partial width would be $2.7\div12.5$~MeV, indeed even larger because of the enhanced phase space compared to the $P_c(4312)$ decay. This is wildly inconsistent with the upper limit of eq. ~\eqref{eq:ld1}. 

To avoid this problem, we have to abandon the assumption that there is a $1/2^-$ $\S\D^*$ bound state, which leads us to Scenario C.
Since for $P_c(4457)$ there are several viable alternatives,  we assume that only $P_c(4440)$ is $\S\D^*$ bound state, and that its quantum numbers are $3/2^-$. We will argue later that there are other reasons to favour this over the alternative Scenarios A and B.

In summary, there is an apparent tension between the dominance of the $P_c(4312)\to\Lc^+\D^{*0}$ decays, and the tight experimental upper limits on $P_c\to \Lc^+\D^0$. This is a problem for Scenarios A and B, but not for Scenario C, where the $\Lc^+\D^0$ decays are small either due to their D-wave nature (for the $3/2^-$ states) or because of heavy quark symmetry (for $P_c(4312)$).

From now on we concentrate on Scenario C. We may estimate the $\Lc\D^*$ partial widths of $P_c(4380)$ and $P_c(4440)$ using equation~\eqref{eq:ld3} and the numbers in Table~\ref{tab:hqs}: 
\begin{align}        \Gamma[P_c^+(4380)\to\Lc^+\D^{*0}]&=2.7\div12.5~\textrm{MeV},\label{eq:ld4}\\
\Gamma[P_c^+(4440)\to\Lc^+\D^{*0}]&=0.9\div4.2~\textrm{MeV}.\label{eq:ld5}
\end{align}
Here we have ignored the effect of the mass differences between $P_c(4312)$, $P_c(4380)$ and $P_c(4440)$. (Phase space would enhance the decays of the heavier states, but this is to some extent mitigated by a corresponding form factor suppression in the matrix element.)

For $P_c(4380)$ we assume the decay width is saturated by $\Lc^+\D^{*0}$ and, from equation~\eqref{eq:G5}, $\Lc\D\pi$. This is because of the tight upper limit on $\jpp$, and the absence of other two-body S-wave decays. (We are assuming that the D-wave decay to $\S\D$ is small.) This suggests a total width
\begin{equation}
    \Gamma[P_c(4380)]=17.7\div 27.5~\textrm{MeV},
\end{equation}
considerably smaller than the measured width \cite{Aaij:2015tga} which, however, is considered to be obsolete~\cite{Aaij:2019vzc}. Other approaches also find a narrower $P_c(4380)$~\cite{Du:2019pij,Du:2021fmf}. Assuming our estimated total width, the $\Lc\D^*$ branching fraction is
\begin{equation}
    {\mathcal B}[P_c(4380)\to\Lc\D^*]=15\%\div 45\%.\label{eq:ld6}
\end{equation}
If neglected decays (notably the $\S\D$ D-wave) are significant, this will be an overestimate.

For $P_c(4440)$, the $\Lc\D^*$ branching fraction is
\begin{equation}
    {\mathcal B}[P_c(4440)\to\Lc\D^*]=3\%\div 45\%,\label{eq:ld7}
\end{equation}
where we have used the measured total width, eq.~\eqref{eq:ld2}, with statistical and systematic uncertainties combined in quadrature.

Given the predicted large  $\Lc^+\D^{*0}$ branching fractions for $P_c(4312)$, $P_c(4380)$ and $P_c(4440)$, searching for these states in $\Lambda_b^0 \to \Lc^+\D^{*0}K^-$ is warranted. We note that this decay has already been observed at LHCb~\cite{Stahl:2018eme}, although an amplitude analysis has not been carried out.

\begin{figure}
    \centering
    \includegraphics[width=0.8\textwidth]{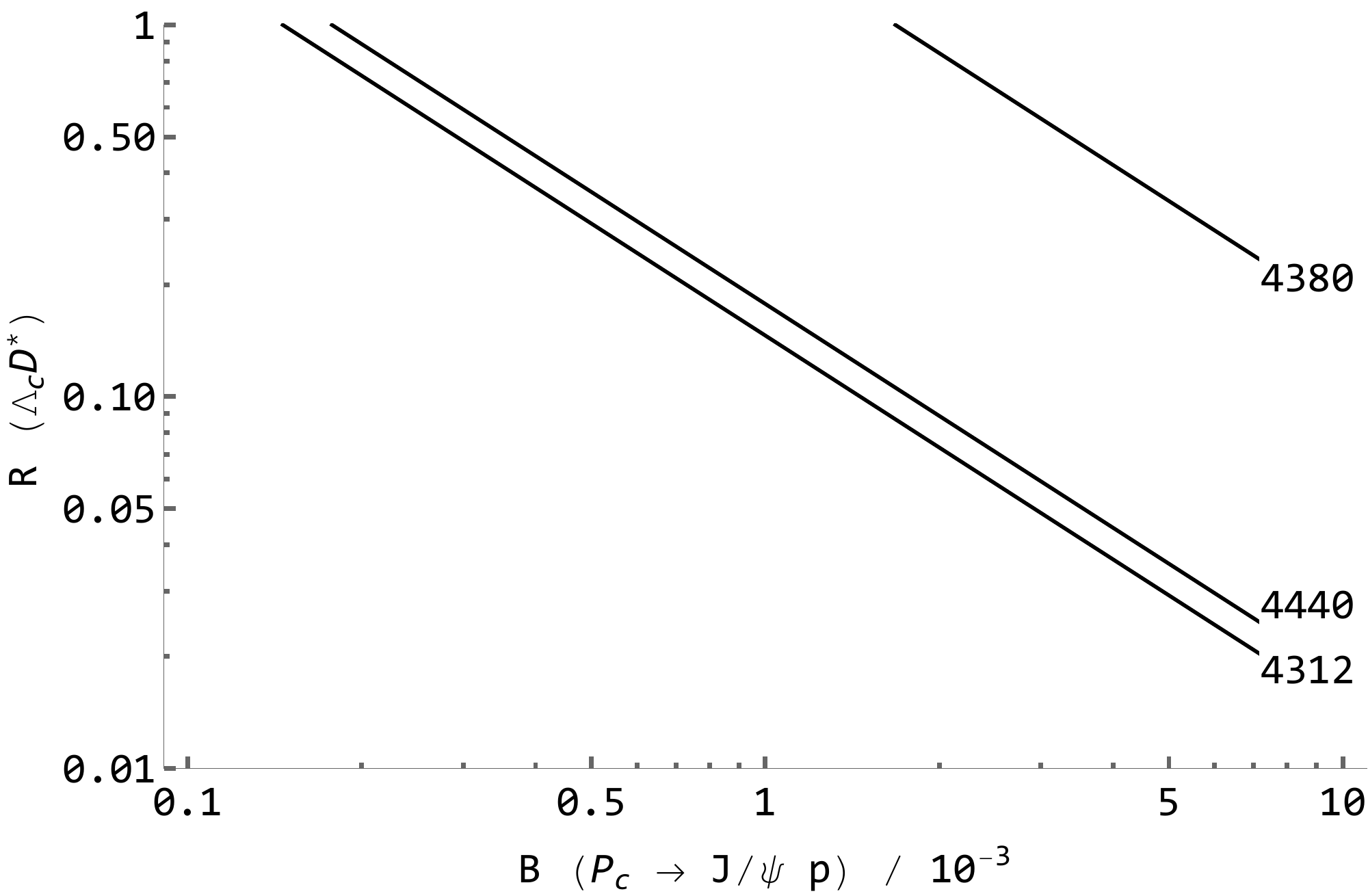}
    \caption{The fit fractions ${\mathcal R}(\Lc^+\D^{*0})$ as a function of ${\mathcal B}[P_c(4312)\to\jpp]$, obtained as described in the text.}
    \label{fig:Rplot}
\end{figure}

To understand the experimental prospects in these channels, we consider now the fit fraction 
\begin{equation}
{\mathcal R (\Lc^+\D^{*0})} = \frac{
{\mathcal B}(\Lambda_b^0 \to P_c^+ K^-) \, 
{\mathcal B}(P_c^+ \to \Lc^+\D^{*0}) } {
{\mathcal B}(\Lambda_b^0 \to \Lc^+\D^{*0}K^-) }.
\label{eq:fflds}
\end{equation}
We may eliminate the unknown production branching fraction ${\mathcal B}(\Lambda_b^0 \to P_c^+ K^-) $ by taking a ratio with equation \eqref{eq:ffjp}, to give
\begin{equation}
    \frac{\mathcal R (\Lc^+\D^{*0})} {\mathcal R (\jpp)} =
    \frac{{\mathcal B}(\Lambda_b^0 \to \jpp\,K^-) }{{\mathcal B}(\Lambda_b^0 \to \Lc^+\D^{*0}K^-) }
    \frac{{\mathcal B}(P_c^+ \to \Lc^+\D^{*0})}{{\mathcal B}(P_c^+ \to \jpp)}.
\end{equation}
Using the experimental numbers in Table~\ref{tab:expt}, and our estimates of ${\mathcal B}(P_c^+ \to \Lc^+\D^{*0})$, we have a relation between the fit fractions $\mathcal R (\Lc^+\D^{*0})$ and the branching fraction ${\mathcal B}(P_c^+ \to \jpp)$, which we plot in Fig.~\ref{fig:Rplot}. Because we are interested in the scale of the numbers rather than their precise values, for illustration we plot the central values only. For $P_c(4380)$, we have used the obsolete value of the fit fraction, equation~\eqref{eq:ff04}.

The message of this plot, when combined with the  upper limit~\eqref{eq:jplimit}, is that the fit fractions ${\mathcal R (\Lc^+\D^{*0})} $ are enormous. Indeed for $P_c(4380)$ the predicted fit fractions are implausibly large. We attribute this to the use of the obsolete value~\eqref{eq:ff04} when constructing the plot. (We encountered a similar problem when interpreting Figure~\ref{fig:bfs}.)

For $P_c(4312)$ and $P_c(4440)$, the fits fractions of order ${\mathcal O}(10\%)$ or larger are still plausible, but are strikingly large when compared to the ${\mathcal O}(1\%)$ fit fractions in the observed $\jpp$ mode (Table~\ref{tab:expt}). It suggests of course that there are strong  prospects for the experimental observation of these states in $\Lambda_b^0 \to \Lc^+\D^{*0}K^-$.

A corollary of this plot is that the branching fraction ${\mathcal B}[P_c(4312)\to\jpp]$ cannot be much less than the current bound in equation~\eqref{eq:jplimit}, otherwise the fit fractions ${\mathcal R}(\Lc^+\D^{*0})$ become implausibly (or impossibly) large. We arrived at the same conclusion when interpreting Fig.~\ref{fig:bfs}. As discussed there, this also implies that observation of $P_c$ states in photoproduction requires not much more sensitivity than that of the $J/\psi${\textit{-007}} experiment.

\section{$\S\*\D$ decays}
\label{sec:sd}

In molecular models, couplings of the type $\S\*\D\*\to \S\*\D\*$ are responsible for binding. The same couplings will also lead to decays into $\S\*\D$ final states, where phase space is available. The $P_c(4312)$ is too light, but $P_c(4380)$ could decay into $\S\D$ (in D-wave), and $P_c(4440)$ could decay into both $\S\D$ (in D-wave) and $\S^*\D$ (in S-wave and D-wave), where we assume the $3/2^-$ quantum numbers of  Scenario C. We further assume that S-wave decays dominate over D-waves, in which case the most prominent decay will be $P_c(4440)\to\S^*\D$. Note that having accounted for $\Lc^+\D^{*0}$ and $\Lc\D^*\pi$ decays, in equations~\eqref{eq:G6} and \eqref{eq:ld5}, respectively, the measured width~\eqref{eq:G2} still allows for a considerable $\S^*\D$ mode. We also note that various models find substantial
$\S^*\D$ decays, for example refs~\cite{Lin:2017mtz,He:2019rva}.

The experimental prospects for observing $P_c(4380)$ or $P_c(4440)$ in $\Lambda_b^0 \to \S\*\D K^-$ depend on the fit fraction 
\begin{equation}
{\mathcal R (\S\*\D)} = \frac{
{\mathcal B}(\Lambda_b^0 \to P_c^+ K^-) \, 
{\mathcal B}(P_c^+ \to \S\*\D) } {
{\mathcal B}(\Lambda_b^0 \to \S\*\D K^-) }.\label{eq:ffsd}
\end{equation}
Taking a ratio with equation \eqref{eq:ffjp}, we get a relation between $\mathcal R (\S\*\D)$ and $\mathcal R (\jpp)$,
\begin{equation}
    \frac{\mathcal R (\S\*\D)} {\mathcal R (\jpp)} =
    \frac{{\mathcal B}(\Lambda_b^0 \to \jpp\,K^-) }{{\mathcal B}(\Lambda_b^0 \to \S\*\D K^-) }
    \frac{{\mathcal B}(P_c^+ \to \S\*\D)}{{\mathcal B}(P_c^+ \to \jpp)}.
    \label{eq:sdratio}
\end{equation}

As far as we know, ${\mathcal B}(\Lambda_b^0 \to \S\*\D K^-) $ has not been measured, but since the tree-level contributions to this mode are colour suppressed (Fig.~\ref{fig:tree}), we expect it to be comparable to ${\mathcal B}(\Lambda_b^0 \to \jpp\,K^-)$, as noted in equation~\eqref{eq:colsup}. In this case, we expect the first ratio on the right-hand side of equation~\eqref{eq:sdratio} to be a number of order 1. Conversely, because of the stringent upper limit~\eqref{eq:jplimit} on $P_c^+\to \jpp$, we expect the second ratio on the right-hand side to be large, at least for $P_c(4440)\to\S^*\D$, where the S-wave is expected to be prominent. We conclude that for $P_c(4440)^+$, the fit fraction $\mathcal R (\S^*\D)$ ought to be large in comparison to the measured fit fraction $\mathcal R (\jpp)$, implying strong experimental prospects. Depending on the magnitude of the D-wave decays, by a similar argument we may also expect considerable (albeit smaller) fit fractions $\mathcal R (\S\D)$ for $P_c(4380)$ and $P_c(4440)$.

This is another example of the general phenomenon discussed in Section~\ref{sec:jpp}. Where tree-level three-body decays are suppressed, two-body fit fractions can be large.

Finally, we remark that, assuming the $P_c$ states are isospin 1/2, the different charge modes $\Lambda_b^0 \to \S^{(*)+}\D^0 K^-$ and $\Lambda_b^0 \to \S^{(*)++} D^- K^-$ have relative rates $1:2$. Deviations from this would be an indication of isospin mixing \cite{Burns:2015dwa}.

\section{Partner states}
\label{sec:partners}

Given the existence of states near the thresholds for $\S\D$, $\S^*\D$ and $\S\D^*$, the apparent absence of states near $\S^*\D^*$ threshold is conspicuous. In this section we show that this is a problem for Scenarios A and B, but not for Scenario C.

Interactions among $\S\*\D\*$ channels are constrained by heavy-quark symmetry. This applies both to models based on meson exchange, and also effective field theory approaches where the long-range contribution to the potential is due to pion-exchange, and the short-range part is modelled via contact terms that are fit to data. 
The pattern of binding in such models can be understood qualitatively with reference to the simplest approach, where the potentials are due to S-wave contact terms only (no meson exchange). The elastic potentials, from  ref.~\cite{Liu:2018zzu}, are given in Table~\ref{tab:elastic}.

\begin{table}[b]
    \centering
    \begin{tabularx}{\textwidth}{XXXXX}
\hline
&$\S\D$&$\S^*\D$&$\S\D^*$&$\S^*\D^*$\\
\hline
$1/2^-$&$C_a$&&$C_a-\frac 4 3 C_b$&$C_a-\frac 5 3 C_b$\\
$3/2^-$& &$C_a$&$C_a+\frac 2 3 C_b$&$C_a-\frac 2 3 C_b$\\
$5/2^-$&&&&$C_a+C_b$\\
\hline
    \end{tabularx}
    \caption{S-wave elastic potentials for $\S\*\D\*$ molecules constrained by heavy quark symmetry.}
    \label{tab:elastic}
\end{table}

Clearly, binding in $\S\D$ and $\S^*\D$ requires $C_a<0$. The potentials in these channels are identical, so apart from small differences due to their masses, and coupled-channel effects, binding in one channel implies binding in the other. Thus a model that accounts for $P_c(4312)$ inevitably implies a partner state $P_c(4380)$. 

To get binding in both of the $\S\D^*$ channels, and thus account for both $P_c(4440)$ and $P_c(4457)$, requires $C_a$ to be large enough (in magnitude) compared to $C_b$. Scenarios A and B are then distinguished by having $C_b>0$ and $C_b<0$, respectively.

With $C_a$ large (in magnitude) compared to $C_b$, on the basis of the potentials in Table~\ref{tab:elastic} it seems likely that all three of the $\S^*\D^*$ states ($1/2^-$, $3/2^-$ and $5/2^-$) will also bind, and that expectation is confirmed by calculation~\cite{Liu:2019tjn}. This immediately raises the question of why there  is apparently no evidence for such states in the experimental data. The absence of the $5/2^-$ state may be understood from its lack of S-wave coupling to $\jpp$; it decays in D-wave, which may be expected to be suppressed.  However the missing $1/2^-$ and $3/2^-$ states are more problematic.

It is possible that Scenarios A and B may be rescued with the idea that the extra $\S^*\D^*$ states do exist, but are just not seen in $\Lambda_b\to J/\psi\, p \,K^-$ because of suppressed production or decays. But in the model we propose \cite{inprep} the situation is precisely the opposite. From heavy quark symmetry, the production and decay of the $3/2^-$ $\S^*\D^*$ state, in particular, would be enhanced compared to other observed states, so if there is binding in this channel, it ought to be visible as a prominent peak in $\Lambda_b\to J/\psi\, p \,K^-$. Similarly, the $1/2^-$ $\S^*\D^*$ state would be expected to decay prominently $\Lc\D$; there is no obvious indication of structure in the $\Lambda_b^0\to \Lc^+\D^0 K^-$ spectrum \cite{Piucci:2019vsk}, but  more detailed analysis with more statistics could be revealing.

The inevitability of binding in all three $\S^*\D^*$ channels ($1/2^-$, $3/2^-$ and $5/2^-$) is a problem that is intrinsic to Scenarios A and B. As noted, to get  binding in both $\S\D^*$ channels ($1/2^-$ and $3/2^-$) implies that $C_a$ must be large in magnitude compared to $C_b$, which in turn renders all of the $\S^*\D^*$ channels sufficiently attractive to cause binding. 

It all plays out very differently if only one of the two $\S\D^*$ were required to bind. In this case it is no longer necessary to have $C_a$ large (in magnitude) in comparison to $C_b$, with the consequence that binding is not automatic in all channels. Indeed, this scenario was already considered after the initial LHCb paper, when the experimental data suggested one $\S\D^*$ state, not two~\cite{Liu:2018zzu}. 

If only one of the  $\S\D^*$ channels binds, it is natural to associate this with  $P_c(4440)$, since this is unambiguously below $\S\D^*$ threshold. As discussed previously, there are several viable alternative scenarios for  $P_c(4457)$, because its mass is consistent with $\S\D^*$ and $\Lc(2595)\D$ thresholds. The next question is whether $P_c(4440)$ has $1/2^-$ or $3/2^-$ quantum numbers. In Section~\ref{sec:ld} we argued in favour of $3/2^-$ quantum numbers (what we called Scenario C), noting that the tight upper limits on $\Lc^+\D^0$ decays are not consistent with a $1/2^-$  $\S\D^*$ state. We will now show that the pattern of binding also argues in favour of the $3/2^-$ assignment.

With reference to Table~\ref{tab:elastic}, to achieve  $\S\D^*$ binding in $3/2^-$, but not $1/2^-$, implies $C_b<0$. This in turn suggests that out of the $\S^*\D^*$ states, there is binding in $5/2^-$, but not $1/2^-$ or $3/2^-$, and indeed this expectation is borne out in explicit calculation~\cite{Liu:2018zzu}. This spectrum of states is nicely consistent with the absence of $\S^*\D^*$ states in the data. The $1/2^-$ and $3/2^-$ states simply don't exist, and the $5/2^-$ state is not visible  because its decay is suppressed. (In our model~\cite{inprep} we find that not only the decay, but also the production,  of the $5/2^-$ state is suppressed.) 

The alternative scenario does not work so well. If instead we assume  $\S\D^*$ binding in $1/2^-$ (but not $3/2^-$), we need $C_b>0$, which implies $\S^*\D^*$ binding in $1/2^-$ and $3/2^-$ (but not $5/2^-$). In this scenario one needs an explanation for why the $1/2^-$ and $3/2^-$  $\S^*\D^*$ states are not visible in $\Lambda_b\to J/\psi\, p \,K^-$. (Note that some authors do argue in favour of the $C_b>0$ pattern, for other reasons~\cite{Chen:2021cfl,Meng:2019ilv}.)

Another reason to prefer binding $\S\D^*$ in $3/2^-$ is that it is also consistent with  expectations from one-pion exchange. In Table~\ref{tab:ope} we show the central potentials due to pion-exchange obtained from the quark model~\cite{Liu:2018zzu,Burns:2019iih}, where $C(r)$ behaves as a Yukawa function (with positive sign) for large $r$ (i.e. it is repulsive at large $r$). It follows that the attractive channels are those in which $C(r)$ comes with a negative coefficient, namely $\S\D^*$ in $3/2^-$ and $\S^*\D^*$ in $5/2^-$; this is consistent with the pattern of binding in our Scenario C. 

Of course the arguments about attraction and repulsion in one-pion exchange are admittedly simplistic, particularly because they ignore the tensor potential, which is known to be important for binding. Nevertheless, explicit calculation including the tensor terms confirms that  the channels that bind most easily (requiring the smallest form factor cutoff) are those identified above ~\cite{Liu:2018zzu,Burns:2019iih,Valderrama:2019chc}. The pattern also applies to models where the potential is due to exchange of not only pions, but also other light mesons~\cite{Liu:2019zvb,Peng:2020xrf}.

\begin{table}
    \centering
    \begin{tabularx}{\textwidth}{XXXXX}
\hline
&$\S\D$&$\S^*\D$&$\S\D^*$&$\S^*\D^*$\\
\hline
$1/2^-$&0&&$4 C(r)$&$5 C(r)$\\
$3/2^-$& &0&$-2C(r)$&$2C(r)$\\
$5/2^-$&&&&$-3C(r)$\\
\hline
    \end{tabularx}
    \caption{S-wave elastic potentials for $\S\*\D\*$ molecules from one-pion exchange in the quark model, where $C(r)$ is the potential defined in ref.~\cite{Burns:2019iih}.}
    \label{tab:ope}
\end{table}
(Notice that the pion exchange potentials in Table~\ref{tab:ope} follow the same pattern as the ``$C_b$'' contact terms in Table~\ref{tab:elastic}, under the replacement $C_b\to -3 C(r)$. This is ultimately because both contributions are proportional to an operator of the form $\mathbf S_1\cdot \mathbf S_2$, where $\mathbf S_1$ and $\mathbf S_2$ are vectors acting on the spin degrees of freedom of the $\S\*$ and $\D\*$, respectively.)

\section{Conclusions}\label{sec:conclusions}
\label{sec:conc}

We have shown that, with minimal theoretical assumptions, current experimental data on $\Lambda_b$ decays and photoproduction imply constraints on the quantum numbers of $P_c$ states, and lead to predictions for their decay  branching fractions, and fit fractions in $\Lambda_b^0$ decays to $\eta_c p K^-$, $\Lc^+\D^{*0}K^-$, and $\S\*\D K^-$.

Unlike the usual Scenarios A and B, where $P_c(4440)$ and $P_c(4457)$ are both $\S\D^*$ bound states, in our Scenario C only $P_c(4440)$ is a bound state, and its quantum numbers are $3/2^-$. Because the $P_c(4457)$
 mass coincides with both $\S\D^*$ and $\Lc(2595)\D$ thresholds, there are several viable alternatives for this state, which we explore in future work~\cite{inprep}. The cusp interpretation of $P_c(4457)$ is favoured by ref.~\cite{Kuang:2020bnk}.
 
One of the advantages of Scenario C is that, unlike Scenarios A and B, it is consistent with the striking observation that the $P_c$ states hardly decay to $\Lc^+\D^0$, but apparently decay prominently to $\Lc^+\D^{*0}$. The suppression of  $\Lc^+\D^0$ is natural for $P_c(4312)$, where it is forbidden by heavy-quark symmetry, and for $P_c(4380)$ and $P_c(4440)$ which, as $3/2^-$ states, decay to $\Lc^+\D^0$ in D-wave. The problem with Scenarios A and B is that each has a $1/2^-$ $\S\D^*$ state which, from heavy quark symmetry,   should decay prominently to $\Lc^+\D^0$, in conflict with experimental data.

Another issue with Scenarios A and B is that they imply the existence of $\S^*\D^*$ partner states, with quantum numbers $1/2^-$, $3/2^-$ and $5/2^-$, which are not apparent in $\Lambda_b\to\jpp\, K^-$ data. This is an inevitable consequence of fixing the potential parameters in order to generate two  $\S\D^*$ bound states, with both $1/2^-$ and $3/2^-$ quantum numbers. Our Scenario~C naturally avoids this issue, as we only require a single $\S\D^*$ state with $3/2^-$ quantum numbers. In this case the potentials are such that the $1/2^-$ and $3/2^-$ $\S^*\D^*$ partners do not bind, and while the $5/2^-$ $\S^*\D^*$ partner does bind,  its absence in experiment can be understood as a consequence of D-wave suppression both in decay and, as discussed in our next paper~\cite{inprep}, production.

A further advantage of Scenario C is that it reproduces the pattern of binding expected on the basis of one-pion exchange.

We have given many predictions for the branching fractions and $\Lambda_b^0$ fit fractions of $P_c(4312)$, $P_c(4380)$ and $P_c(4440)$ in various modes. The most striking prediction is for strong experimental signals for all the states in $\Lambda_b^0\to\Lc^+\,\D^{*0}\,K^-$. This is a particularly sharp prediction when juxtaposed against the absence of signals in the closely related decay $\Lambda_b^0\to\Lc^+\,\D^{0}\,K^-$. The decay $\Lambda_b^0\to\Lc^+\,\D^{*0}\,K^-$ has been observed at LHCb~\cite{Stahl:2018eme}, and we suggest an amplitude analysis to test for the existence of $P_c$ states in the $\Lc^+\,\D^{*0}$ spectrum. 

In $\Lambda_b^0\to\eta_c\,p\,K^-$, we predict a substantial $P_c(4312)$ fit fraction, which can be tested in future experimental analyses. (Our prediction is consistent with the current upper bound.) In our Scenario C, we do not expect prominent signals for the other $P_c$ states in $\Lambda_b\to\eta_c\,p\,K^-$. We showed that, by contrast, Scenarios A and B could be distinguished by the relative fit fractions of $P_c(4440)$ and $P_c(4457)$.

We also predict strong experiment signals for $P_c(4440)$ in $\Lambda_b^0 \to \S^*\D K^-$, and smaller signals for both $P_c(4380)$ and $P_c(4440)$ in $\Lambda_b^0 \to \S\D K^-$.

We also pointed out that the branching fractions ${\mathcal B}(P_c^+\to J/\psi\, p)$ cannot be much less than the current upper bound \eqref{eq:jplimit}, otherwise the production branching fractions $\mathcal B (\Lambda_b^0\to P_c^+ K^-)$ and fit fractions ${\mathcal R}(\Lc^+\D^{*0})$ would be implausibly large.
(See Figs.~\ref{fig:bfs} and~\ref{fig:Rplot}.) An interesting consequence is that existing experimental measurements are almost at the level of sensitivity required to observe $P_c$ states in photoproduction.

We want to emphasise that many of the constraints and predictions we have derived are very general in nature, and so if it turns out they are not satisfied by future experimental data, it could be an indication not of the failure of our particular realisation of the molecular model, but of the molecular approach all together. As an example, consider again the observation  (Fig.~\ref{fig:bfs}) that if the upper limit on ${\mathcal B}(P_c^+\to J/\psi\, p)$ is much less than equation~\eqref{eq:jplimit},  it would imply implausibly large production branching fractions $\mathcal B(\Lambda_b^0\to P_c^+K^-)$. In this case some new mechanism would need to be invoked to explain how, contrary to intuition, molecular states are produced with comparable branching fraction to conventional hadrons. Alternatively, since the observation is a direct consequence of the measured fit fractions, one could question whether or not it is appropriate, in the fit fraction, to factorise the production and decay branching fractions $\mathcal B(\Lambda_b^0\to P_c^+K^-)$ and ${\mathcal B}(P_c^+\to J/\psi\, p)$ in the numerator. The factorisation is legitimate in the case that $P_c$ is a resonance, but not if the $P_c$ signal is an effect which is uniquely associated with a particular production mechanism, such as a cusp or triangle singularity. Indeed we gave examples, in the case of the $X(2900)$ states, where the factorisation is clearly not appropriate~\cite{Burns:2020xne}. Many of our manipulations in this paper rely on this factorisation, and so our predictions should be interpreted with this in mind.

\acknowledgments
Swanson's research was supported by the U.S. Department of Energy under contract DE-SC0019232.

\bibliography{bibfile}

\end{document}